\documentclass[MSNbibl,nameyear,dvips]{arxstspdf}
\usepackage{flushend}
\usepackage{stfloats}

\volume{29}
\issue{2}
\pubyear{2014}
\firstpage{252}
\lastpage{253}
\doi{10.1214/14-STS473} 
\referstodoi{10.1214/13-STS457}

\makeatletter

\renewcommand{\citep}[1]{(\cite{#1})}
\makeatother

\begin{document}
\begin{frontmatter}

\vspace*{6pt}\title{Discussion: On Arguments Concerning Statistical Principles}
\runtitle{Discussion}

\begin{aug}
\author[a]{\fnms{D. A. S.}~\snm{Fraser}\corref{}\ead[label=e1]{dfraser@utstat.toronto.edu}}
\runauthor{D. A. S.~Fraser}

\affiliation{University of Toronto}

\address[a]{D. A. S. Fraser is Emeritus Professor,
Department of Statistical Sciences,
University of Toronto,
100 St. George St., Toronto, Ontario M5S 3G3, Canada \printead{e1}.}
\end{aug}


\end{frontmatter}

(i) \textit{Statistical inference after Neyman--Pearson}.
Statistical inference as an alternative to Neyman--Pearson decision
theory has a long history in statistical thinking,
with strong impetus from Fisher's research; see, for example, the
overview in \citet{Fisher:1956}. Some resulting concerns in inference
theory then reached the mathematical statistics community rather
forcefully with \citet{Cox:1958}; this had focus on the two
measuring-instruments example and on uses of conditioning that were compelling.

(ii) \textit{Birnbaum and logical analysis in statistical inference}.
\citet{Birnbaum:1962} introduced notation for the statistical
inference available from an investigation with a model and data. This
gave grounds to analyze how different methods or principles might
influence the statistical inference. As part of this he discussed how
sufficiency, likelihood and conditioning could differentially affect
statistical inference. Much of his discussion centered on the argument
from conditioning and sufficiency to likelihood, but a primary consequence
was the attention attracted to conditioning and its role in inference.
While this interest in conditioning was substantial for those concerned
with the core of statistics, it has more recently been neglected or
overlooked. Indeed, some recent texts, for example, \citet{Rice:2007},
seem not to acknowledge conditioning in inference or even the
measuring-instrument example.

(iii) \textit{Mayo and statistical principles}.
Mayo should be strongly commended for reminding us that the principles
and arguments of statistical inference deserve very serious
consideration and, we might add, could have very serious consequences
\citep{Fraser:2013c}. Her primary focus
is on the argument \citep{Birnbaum:1962} that the principles
sufficiency and conditionality lead to the likelihood principle. This
may not cover some recent aspects of conditioning \citep
{FraserFS:2010}, but should strongly stimulate renewed interest in conditioning.

(iv) \textit{Contemporary inference theory}.
Many statistical models have continuity in how parameter change affects
observable variables or, more specifically, how parameter change
affects coordinate quantile functions, the inverses of the coordinate
distribution functions. This continuity in its global effect is widely
neglected in statistical inference. If this effect on quantile
functions is accepted and
used in the inference procedures, then in wide generality there is a
well-determined conditioning \citep{FraserFS:2010}. And likelihood
analysis then offers an exponential model approximation that is
third-order equivalent to the given model, and this in turn provides
third-order inference for any scalar component parameters of interest.
Thus, the familiar conditioning conflicts are routinely avoided by
acknowledging the important model continuity.

(v) \textit{What is available?}
The conditioning just described leads routinely to $p$-value functions
$p(\psi)$ for any scalar component parameter
$\psi=\psi(\theta)$ of the statistical model. A wealth of
statistical inference methodology then immediately becomes available
from such $p$-value functions. For example, a test for a value $\psi
_0$ is given by the $p$-value $p(\psi_0)$, a confidence interval by
the inverse $(\hat\psi_{\beta/2},\hat\psi_ {1-\beta
/2})=p^{-1}(1-\beta/2,\beta/2)$ of the $p$-value function, and a
median estimate by
the value $p^{-1}(0.5)$. But quite generally the needed $p$-value
functions are not available from a likelihood function alone!

(vi) \textit{What are the implications?}
If continuity is included as an ingredient of many model-data
combinations, then, as we have indicated, likelihood analysis produces
$p$-values and confidence intervals, and these are not available from
the likelihood function alone. This thus demonstrates that with such
continuity-based conditioning the likelihood principle is not a
consequence of sufficiency and conditioning principles. But if we omit
the continuity then we are directly faced with the issue addressed by Mayo.

\section*{Acknowledgments}
Supported in part by the Natural Sciences and Engineering
Research Council of Canada and Senior Scholars Funding at York University.



\end{document}